\documentclass[conference]{IEEEtran}

\usepackage{graphicx}
\usepackage{subcaption}
\usepackage{balance}
\usepackage{comment}
\usepackage{amsmath}

\usepackage{balance}
\usepackage{stfloats}

\usepackage{multirow}
\usepackage[table]{xcolor}

\IEEEoverridecommandlockouts
\begin{document}

\title{Sparsity-Aware Hardware-Software Co-Design of Spiking Neural Networks: An Overview}

	\author{\IEEEauthorblockN{Ilkin Aliyev\IEEEauthorrefmark{1}, Kama Svoboda\IEEEauthorrefmark{1}, Tosiron Adegbija\IEEEauthorrefmark{1}, Jean-Marc Fellous\IEEEauthorrefmark{2}}
	\IEEEauthorblockA{\IEEEauthorrefmark{1}Department of Electrical \& Computer Engineering \\ \IEEEauthorrefmark{2}Departments of Psychology and Biomedical Engineering \\ University of Arizona, Tucson, AZ, USA \\
	Email: \{ilkina,ksvoboda,tosiron,fellous\}@arizona.edu}}

	
\maketitle
	
\begin{abstract}
Spiking Neural Networks (SNNs) are inspired by the sparse and event-driven nature of biological neural processing, and offer the potential for ultra-low-power artificial intelligence. However, realizing their efficiency benefits requires specialized hardware and a co-design approach that effectively leverages sparsity. We explore the hardware-software co-design of sparse SNNs, examining how sparsity representation, hardware architectures, and training techniques influence hardware efficiency. We analyze the impact of static and dynamic sparsity, discuss the implications of different neuron models and encoding schemes, and investigate the need for adaptability in hardware designs.  Our work aims to illuminate the path towards embedded neuromorphic systems that fully exploit the computational advantages of sparse SNNs.
\end{abstract}



\section{Introduction}

Energy-efficient and high-performance computing architectures have become more essential than ever in the era of pervasive machine learning (ML) and artificial intelligence (AI). Spiking Neural Networks (SNNs), which mimic the event-driven communication of biological neurons, hold the promise of surpassing the energy efficiency of conventional Artificial Neural Networks (ANNs) \cite{tavanaei19_deepSNN}. An important reason for the potential efficiency of SNNs is that they exploit the inherent sparsity observed in biological neural systems, characterized by \textit{sparse coding} \cite{foldiak03_sparseCoding,spanne15_sparseCoding} and \textit{sparse connectivity} \cite{faghihi22_sparseConnectivity,eavani15_sparseConnect}, and computation using partial synchrony instead of firing rate \cite{brette12_computingSynchrony}. This sparsity translates directly into potential computational savings in hardware implementations, especially if the sparsity is explicitly exploited \cite{yin22_snn_sparsity_accel,aliyev24_pulse}.

In sparse coding, only a fraction of neurons are activated at a time. As a result, hardware designs that explicitly exploit sparse coding in SNN models can conserve energy by powering down inactive neurons, thus only consuming power for processing active signals on demand. This selective activation aligns well with event-driven processing, where computations are performed only when events (spikes) occur, reducing the overall energy consumption. Moreover, sparse connectivity implies that each neuron is connected to only a subset of other neurons, rather than a fully connected network. This reduces the complexity of the inter-neuronal communication infrastructure required. For an SNN hardware accelerator, this translates to fewer necessary connections and routing paths, which can simplify the accelerator's physical layout and reduce the energy costs associated with data transfer and storage. Furthermore, the high reliability of digital hardware (compared to biological neurons) makes it possible to increase the sparsity of the computations beyond what is observed in the brain to potentially achieve even more energy efficiency.

However, translating this theoretical efficiency into tangible gains on real-world hardware remains a critical challenge. Specialized hardware platforms that explicitly exploit the characteristics of SNN workloads are necessary to reap the full benefits of sparse SNN computations \cite{yin22_snn_sparsity_accel,aliyev24_pulse}. As such, the co-design of hardware and software holds the key to unlocking the energy-saving potential of SNNs. Algorithms and models must be tailored to work in synergy with hardware architectures optimized to handle the unique computational characteristics of sparse SNNs.

Key considerations in this co-design process involve how sparsity is represented and how it interacts with the underlying hardware. Sparse operations in SNN computations often require different approaches for hardware acceleration than dense operations. Additionally, model configurations, such as the synaptic connectivity patterns, neuron models, encoding schemes, and the balance between different kinds of sparsity, including static and dynamic sparsity, can have profound impacts on hardware efficiency. For example, static sparsity, which refers to a fixed pattern of zero-valued weights in the SNN model, allows for predetermined optimizations like memory compression and skipping of computations with zero weights. On the other hand, dynamic sparsity, referring to the temporal event-based neuron activations, offers potential for further efficiency, but requires flexible hardware to handle variable, irregular, and unpredictable computational loads.

This paper provides an overview of the multifaceted field of hardware-software co-design for sparse SNNs, emphasizing the critical role of sparsity in achieving energy-efficient neuromorphic computing. We investigate the dynamic nature of sparsity, exploring its dependency on various factors such as training hyperparameters, neuron models, and input encoding methods. Through empirical analysis and detailed exploration, we quantify the impact of these factors on sparsity, offering valuable insights for optimizing SNNs for hardware implementation. We also address the challenges inherent in hardware-software co-design of SNNs, highlighting the need for specialized hardware architectures and sparsity-aware training techniques. Furthermore, we survey existing hardware architectures and techniques specifically designed to exploit sparsity, showcasing the potential for significant performance and energy efficiency gains in SNN accelerators. By bridging the gap between theoretical potential and practical implementation, this paper aims to contribute to the advancement of neuromorphic computing and pave the way for a new generation of energy-efficient AI applications.

\section{Background on Spiking Neural Networks}

\begin{figure}
		\centering
		\includegraphics[width=0.25\textwidth]{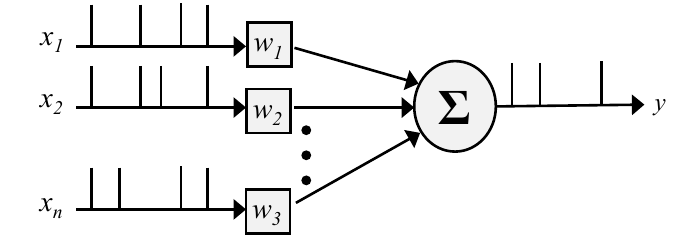}
		\caption{SNN neurons integrate incoming spikes $x$ with corresponding synaptic weights $w$ to generate output spikes $y$ every time the integrated membrane potential reaches a threshold.}
		\label{fig:neuron}
\end{figure}

While both SNNs and ANNs ultimately map input patterns to outputs, their computational models differ significantly.  ANNs rely on continuous-valued activation functions, whereas SNNs utilize discrete binary spikes within the temporal domain to represent information \cite{ponulak11_snnIntro}.  As depicted in Figure \ref{fig:neuron}, using the integrate-and-fire neuron model \cite{burkitt06review} as an example, SNN neurons accumulate incoming spikes, integrating their weighted influence over time. A neuron fires an output spike only when its membrane potential surpasses a defined threshold. In this section, we present a brief overview of SNNs underpinned by the importance of sparsity as a core feature.

\subsubsection{Neuron models} At the heart of an SNN lies the individual neuron model and its synapses which determine the network's learning dynamics. Simple neuron models like the Leaky Integrate-and-Fire (LIF) \cite{lansky06_lif_neuron} mimic the thresholding behavior of neurons, i.e., spikes are generated when their membrane potential exceeds a threshold. More complex models, such as the Hodgkin-Huxley \cite{nelson95_hodgkin}, do not have an explicit threshold but detail the dynamics of membrane ion channels for greater biological realism and introduce computational overhead.  The choice of neuron model profoundly impacts the efficiency obtained from network sparsity, learning dynamics, and the suitability for different hardware implementations.

\subsubsection{Spatiotemporal dynamics}  SNNs fundamentally differ from traditional ANNs in their approach to information processing. While both utilize the activation patterns of neurons to encode information, SNNs introduce the precise timing of neuronal spikes as an additional dimension \cite{she21_snnSpatioTemporal}. This timing allows neurons to convey information through single spikes, bursts, or complex temporal patterns.  Learning mechanisms like Spike Timing Dependent Plasticity (STDP) \cite{serrano13_stdp}, which modify synaptic strengths based on the relative timing between pre- and post-synaptic spikes, enable SNNs to learn both spatial and temporal patterns. This unique capability positions SNNs for applications in sequence recognition, temporal prediction, and adaptive behavior within dynamic environments.  Furthermore, the inherent temporal sparsity of SNNs, where neurons only fire when necessary, contributes significantly to their energy efficiency \cite{han20_snnEnergyEfficiency}. 

\subsubsection{Learning in SNNs} Although backpropagation \cite{rojas96_backpropagation} has become a workhorse for effectively training ANNs in practice, training SNNs presents unique challenges due to the non-differentiability of spike-based signals. Several training methods address this, including \textit{ANN-to-SNN conversion} \cite{bu23_ann_snn}, where a conventional ANN is trained and then converted to an SNN, potentially sacrificing accuracy and efficiency gains.  Unsupervised methods utilize STDP \cite{serrano13_stdp}, but often suffer from slow convergence, high sensitivity to noise and high sensitivity to parameter setting. More recently, supervised learning with \textit{surrogate gradients} \cite{neftci19_surrogateGradients} has shown promise by using differentiable surrogate functions during backpropagation-like training, allowing optimization of SNNs for both accuracy and hardware efficiency. 

\subsubsection{Input encoding}  Input encoding in SNNs, the translation of real-world data into spikes, significantly affects information representation, network sparsity, robustness to noise, and hardware efficiency. Different encoding methods offer distinct trade-offs. For example, rate coding \cite{rathi20_rateCoding} encodes information in the average firing rate over time, offering high performance in deep networks (e.g., VGG9, VGG11) \cite{kang2020asie}, but often at the cost of reduced sparsity due to the high spike rate. Temporal coding \cite{zhou21_temporalCoding}, conversely, focuses on the precise timing of spikes or patterns of spikes within short time frames \cite{fellous04_spikePatterns}. While generally sparser than rate coding, it can sometimes lead to lower accuracy, though methods like \textit{time-to-first-spike (TTFS)} have achieved high accuracy in certain applications \cite{sommer2022efficient,lew2022time}. Delta encoding \cite{yarga22_deltaEncoding} strikes a balance by using the temporal change of input features to generate spikes, offering a compromise between sparsity and accuracy. Radix encoding \cite{wang2022efficient} aims for ultra-short spike trains, achieving high accuracy with few time steps, but may require specialized hardware. Direct coding \cite{wu19_snn_training} bypasses explicit input encoding, allowing the training algorithm to learn the optimal mapping of input data to spiking patterns. The choice of encoding scheme depends on various factors, including input data characteristics, neuron models, and the target application.

\subsubsection{Applications}  SNNs are well-suited for tasks where temporal dynamics and efficient processing are vital (e.g., machine learning implementations on resource-constrained devices). They are particularly well-suited for processing data from event-based sensors (such as neuromorphic vision sensors or dynamic audio sensors) \cite{singh21_gestureSNN}, where the sensor output aligns naturally with the sparse, spike-based communication in SNNs. This enables low-power, real-time processing in resource-constrained edge computing systems. SNNs also show promise in embedded pattern recognition tasks \cite{kim18_SNNPatternRecognition} where stringent power constraints must be adhered to.  Their ability to learn temporal patterns makes them applicable to tasks such as gesture recognition \cite{amir17_dvsGesture}, anomaly detection in time-series data \cite{yusob18_SNNanomalyDetection}, or adaptive control systems \cite{nichols12_snnRobot}. Additionally, the biological plausibility of SNNs opens up opportunities for computational and experimental neuroscience research, enabling the modeling of specific computations implemented by a given brain region and the investigation of learning and memory mechanisms.

\section{Neurobiological Foundations of Sparsity}

Neuroscience research reveals that sparsity may be fundamental to the brain's organization and function, influencing storage \cite{foldiak03_sparseCoding}, energy consumption \cite{yu17_sparsityEnergy}, robustness to noise \cite{bricken23_sparseNoise}, and processing efficiency \cite{schweighofer01_sparseCoding}. This sparsity manifests in various ways:

\subsubsection{Sparse neural coding} The brain employs a sparse distributed coding scheme, where only a small subset of neurons are active in response to specific stimuli or tasks, enhancing energy efficiency and robustness to noise \cite{xu09_sparseCoding,li16_robust_sparsity}.

\subsubsection{Structural sparsity} The brain exhibits a high degree of sparse connectivity---i.e., neurons form connections with only a fraction of other neurons \cite{eavani15_sparseConnect}. This minimizes metabolic wiring costs and promotes modular and specialized subnetworks for efficient processing. 

\subsubsection{Sparsity, plasticity, and learning} Sparsity interacts dynamically with learning mechanisms such as STDP \cite{hassall18_sparsity_learning}, allowing for flexible synaptic modifications and synaptic pruning, which refines network representation during development and learning \cite{gerum20_sparsityPruning}.

\subsubsection{Computational models of sparsity} Theoretical models suggest that sparsity enhances brain computing power by reducing redundancy and facilitating pattern separation \cite{herbert22_sparsityModels}, aiding in classification tasks \cite{harris19_sparsityModel}. 

Understanding the biological basis of sparsity is crucial for developing neuromorphic computing systems that aim to mimic the brain's efficiency and low power consumption. Insights from biological sparsity can inspire the design of algorithms, hardware optimizations, and plasticity mechanisms for more efficient AI in resource-constrained systems.

\section{Understanding the Dynamics of Sparsity in Practical SNNs}

The inherent sparsity of SNNs is key to their energy efficiency. Sparsity is a dynamic property influenced by various factors like the network's training algorithms, neuron models, input encoding methods, and dataset characteristics. This section explores the impact of neuron models, their hyperparameters, and encoding methods on sparsity.

\subsection{Sparsity in the LIF and Lapicque neuron models}

To examine the sparsity characteristics of neuron models, we consider two simple models: \textit{Lapicque} \cite{brunel07_lapicque} and \textit{leaky integrate-and-fire (LIF)} \cite{lansky06_lif_neuron}. The Lapicque model, introduced in 1907, represents a neuron as a single point with a membrane potential that evolves in response to incoming inputs ($I(t)$). If the membrane potential ($u_j(t)$) exceeds a threshold ($\theta$), the neuron fires a spike and resets to its resting value ($u_{rest}$):

\begin{equation}
\frac{du_j(t)}{dt} = 
\begin{cases}
    I(t) & \text{if } u_j(t) < \theta \\
    u_{rest} & \text{if } u_j(t) \geq \theta
\end{cases}
\end{equation}

Due to its simplicity, sparsity emerges naturally in the Lapicque model. If the input is insufficient to push the membrane potential above the threshold, the neuron remains silent. Sparsity in this model is primarily determined by the distribution of input weights and the chosen threshold value. However, its lack of temporal dynamics limits the complexity of sparsity patterns it can exhibit.

The LIF model extends the Lapicque model by introducing a ``leak" term, simulating the gradual decay of the membrane potential towards its resting state. The interplay between input strength, the membrane potential's leak term, and the firing threshold governs the neuron's spiking behavior. The leak's time constant influences how quickly the neuron ``forgets" previous inputs, impacting sparsity. A shorter time constant leads to a more rapid decay of the membrane potential in the absence of new inputs and effectively increases the amount of input current required to reach the firing threshold, thereby leading to sparser activity. The LIF neuron's characteristics can be expressed as: 

\begin{equation} \label{ref:lif1}
    u_{j}[t+1] = \beta \cdot u_{j}[t] + \sum_{i} w_{ij} \cdot s_{i}[t] - s_{j}[t] \cdot \theta
\end{equation}

\begin{equation}\label{ref:lif2}
s_{j}[t] = 
\begin{cases} 
1, & \text{if } u_{j}[t] > \theta \\
0, & \text{otherwise}
\end{cases}
\end{equation}

\noindent where $\beta$ (decay factor) controls the membrane potential decay rate, and impacts how the previous potential $u_{j}[t]$ affects the current potential $u_{j}[t+1]$. $\theta$ represents the firing threshold to produce a spike $s_{j}[t]$. A higher $\beta$ and $\theta$ can lead to sparser firing. More complex neuron models can similarly be analyzed based on their configurable parameters.

\subsection{Practical impacts of model hyperparameters on sparsity}

Model hyperparameters can significantly influence SNN sparsity and hardware efficiency. For example, a previous study \cite{aliyev24_surrogateGradients} showed that using the fast sigmoid surrogate gradient function instead of arctan increased sparsity and improved frames per second/watt (FPS/W) by 11\%. Fine-tuning neuronal parameters like decay rate and threshold further reduced latency by 48\% with minimal accuracy loss.

To further examine the LIF and Lapicque neuron models, we performed experiments to explore the impacts of their different hyperparameters on sparsity. These experiments underscore the importance of sparsity-aware hardware-software co-design in the development of SNNs, illustrating the need for carefully balancing the trade-offs between accuracy and sparsity. 

We used snnTorch \cite{eshraghian21_snntorch} to build spiking neuron models and PyTorch to train a convolutional SNN (CSNN) on the Street View House Numbers (SVHN) dataset. We used a VGG-9-based \cite{dong2017learning} CSNN architecture with the structure: 64C3-P1-112C3-P1-192C3-P1-216C3-P1-480C3-P1-504C3-P1-560C3-P1-1064FC-P1-5000FC-P1-Dropout, where $x$C$y$ denotes convolutional layers with $x$ filters of size $y \times y$. Depending on the neuron model employed, P1 represents either the LIF or Lapicque layer. $x$FC is a fully connected layer containing $x$ neurons. Training was performed for 200 epochs. Given prior research on its ability to enhance sparsity \cite{aliyev24_surrogateGradients}, we used the fast sigmoid surrogate gradient function for training the network. Network parameters were updated via the Adam optimizer with an initial learning rate of $5.0 \times 10^{-3}$.

\subsubsection{Beta-threshold exploration} 
\begin{figure}
\centering
  \begin{subfigure}[b]{0.4\textwidth}
    \includegraphics[width=\textwidth]{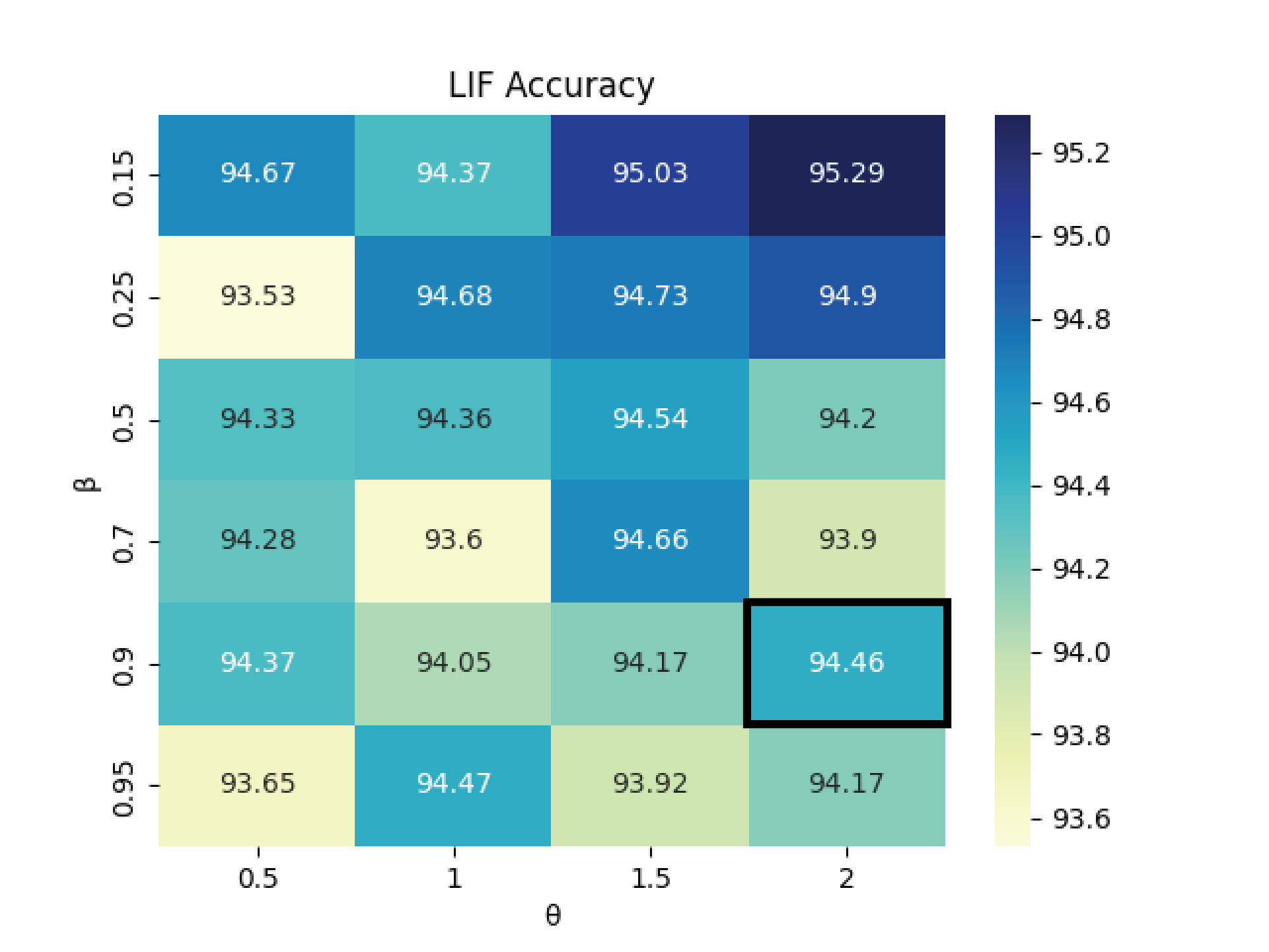}
    \caption{LIF Accuracy}
    \label{fig:LIF_cross_sweep_a}
  \end{subfigure}
 \hfill
  \begin{subfigure}[b]{0.4\textwidth}
    \includegraphics[width=\textwidth]{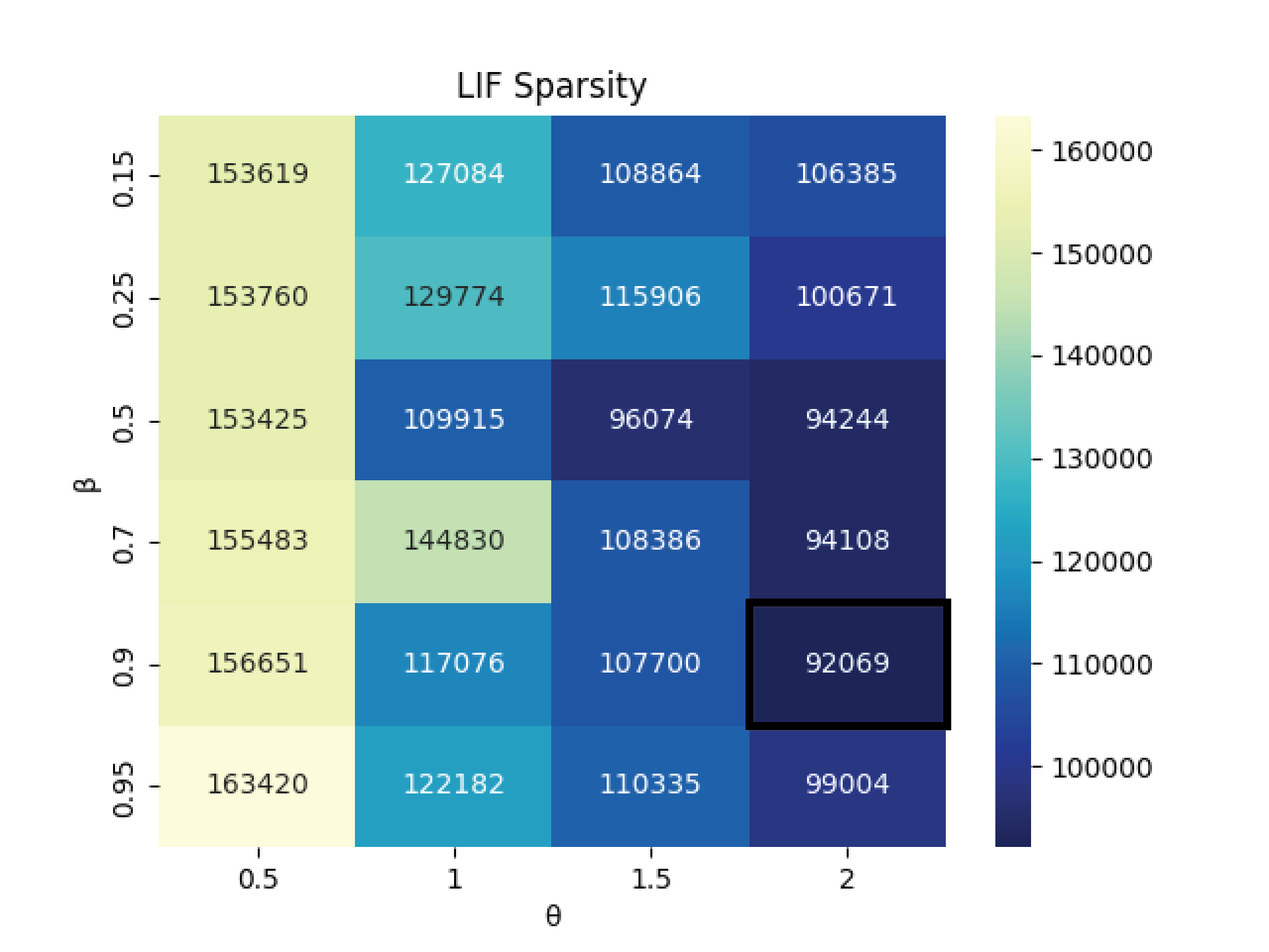}
    \caption{LIF Sparsity}
    \label{fig:LIF_cross_sweep_b}
  \end{subfigure}
  \caption{LIF neuron model cross sweep results for $\beta$ and $\theta$ parameters.}%
  \label{fig:LIF_cross_sweep}
\end{figure}

\begin{figure}
\centering
  \begin{subfigure}[b]{0.4\textwidth}
    \includegraphics[width=\textwidth]{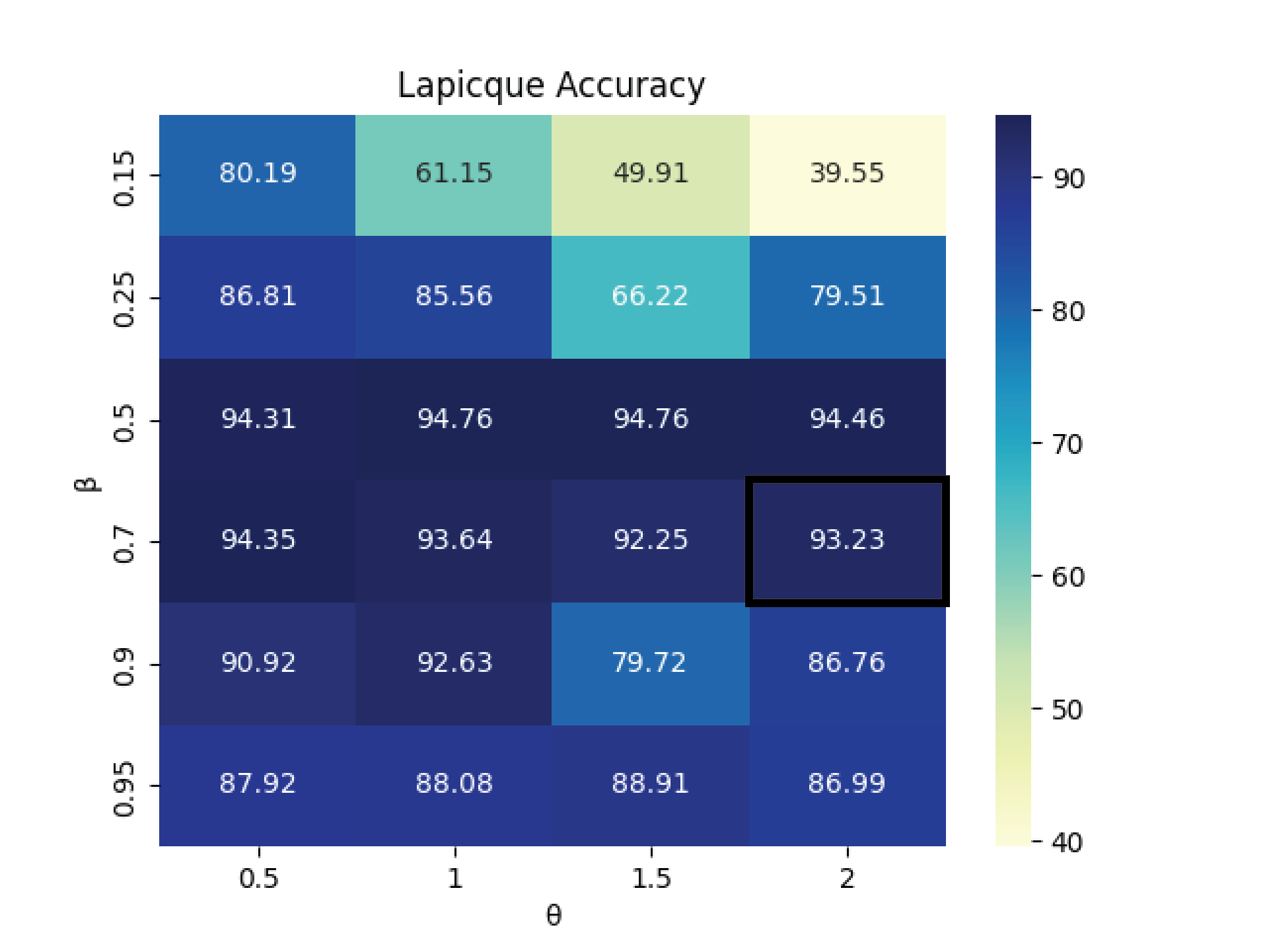}
    \caption{Lapicque Accuracy}
    \label{fig:Lapicque_cross_sweep_a}
  \end{subfigure}
 \hfill
  \begin{subfigure}[b]{0.4\textwidth}
    \includegraphics[width=\textwidth]{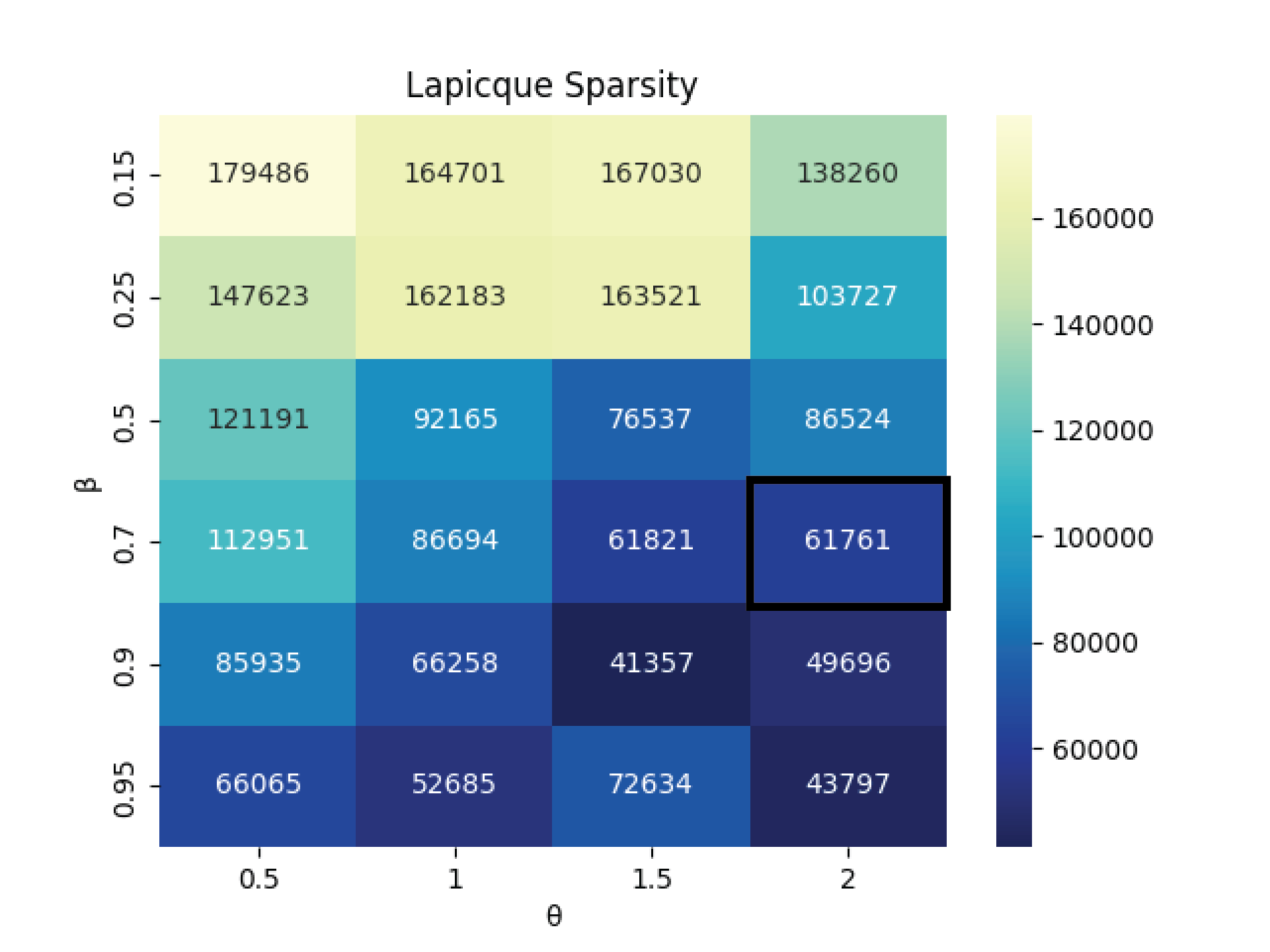}
    \caption{Lapicque Sparsity}
    \label{fig:Lapicque_cross_sweep_b}
  \end{subfigure}
  \caption{Lapicque neuron model exploration for $\beta$ and $\theta$.}%
  \label{fig:Lapicque_cross_sweep}
  \vspace{-10pt}
\end{figure}

We started by performing a detailed exploration of $\beta$ (the leakage factor) and $\theta$ (the firing threshold) for both the LIF and Lapicque neuron models using direct encoding, evaluating accuracy and sparsity. The Lapicque neuron is implemented using RC circuit parameters, with $\beta = e^{-\frac{1}{RC}}$. $R$ (resistance) defaults to 1 and $C$ (capacitance) is inferred based on the value of $\beta$. The results of these explorations can be seen in Figures \ref{fig:LIF_cross_sweep} and \ref{fig:Lapicque_cross_sweep}. 

In Figure \ref{fig:LIF_cross_sweep_a}, we observe that the accuracy using the LIF neuron remains relatively high across various values of $\beta$ and $\theta$, with a maximum accuracy of 95.29\% achieved at $\beta = 0.15$ and $\theta = 2$. Figure \ref{fig:LIF_cross_sweep_b} shows the LIF sparsity, measured by the number of spikes. Here, we see that in general, the sparsity increases (fewer spikes) as $\beta$ and $\theta$ increase. The sparsest configuration, with 92,069 spikes, occurs at $\beta = 0.9$ and $\theta = 2$. These observations suggest that higher thresholds and decay factors encourage more selective neuronal activity, leading to higher sparsity. Combining insights from both figures illustrates the trade-offs between sparsity and accuracy. While high sparsity, and in effect, hardware efficiency, can be achieved by increasing the threshold and decay factor, it may come at a cost to accuracy. As such, a balance must be found where accuracy remains high without significantly compromising on sparsity. In this case, for example, the sparsest configuration ($\beta = 0.9$, $\theta = 2$) might represent a satisfactory balance when hardware efficiency is the priority. This configuration increased sparsity by 13.5\% compared to the best accuracy configuration, with only a 0.83\% decrease in accuracy, indicating that this configuration is efficient in terms of sparsity while maintaining high performance.

To illustrate the diversity in the sparsity of different neuron models and the need for detailed exploration of software and hardware configurations, Figure \ref{fig:Lapicque_cross_sweep} depicts a similar exploration for the Lapicque neuron. In Figure \ref{fig:Lapicque_cross_sweep_a}, we observe a notable decline---more so than for the LIF neuron---in accuracy as $\theta$ increases, particularly at lower $\beta$ values. Interestingly, the accuracy remains relatively robust at higher decay factors ($\beta = 0.95$), even with an increase in threshold, maintaining a minimum accuracy of 86.99\%. 

As with the LIF neuron, the sparsity for the Lapicque neuron increases as $\beta$ and $\theta$ increase. In this case, the optimal balance for the Lapicque neuron model was found with $\theta$ at 2.0 and $\beta$ at 0.7, achieving 93.23\% accuracy and 61,761 spikes. This configuration increased sparsity by 33.0\% compared to the best accuracy configuration, with a 1.53\% accuracy loss. These results suggest that while the Lapicque neuron might be more sensitive to changes in $\beta$ and $\theta$ than the LIF neuron, the Lapicque neuron might be more efficient regarding sparsity than the LIF model while maintaining comparable accuracy.

\begin{table}[t!]
\centering
\caption{Accuracy and sparsity for LIF and Lapicque models across different encoding methods.}
\begin{tabular}{ |p{1.2cm}|p{1.2cm}|p{1.2cm}|p{1.2cm}| }
 \hline
 \textbf{Encoding} & \textbf{Model} & \textbf{Accuracy} & \textbf{Spikes} \\
 \hline
 \multirow{2}{*}{Rate} & LIF & 77.99\% & 1,091,195 \\
 \cline{2-4}
                       & Lapicque & 75.88\% & 690,700 \\
 \hline
 \multirow{2}{*}{Delta} & LIF & 38.40\% & 79,969 \\
 \cline{2-4}
                        & Lapicque & 40.53\% & 34,246 \\
 \hline
 \multirow{2}{*}{Direct} & LIF & 94.46\% & 92,069 \\
 \cline{2-4}
                         & Lapicque & 93.23\% & 61,761 \\
 \hline
\end{tabular}
\label{table:encoding}
\end{table}

\subsubsection{Impacts of encoding methods} 
To investigate how different encoding approaches affect accuracy and sparsity, we used the optimal $\beta$ and $\theta$ (highlighted in Figures \ref{fig:LIF_cross_sweep} and \ref{fig:Lapicque_cross_sweep}) for the LIF and Lapicque neurons, respectively, and trained the model using each encoding methods. Table \ref{table:encoding} presents a comparative analysis of the accuracy and sparsity of LIF and Lapicque neuron models across three encoding methods: rate encoding, delta encoding, and direct encoding. 

The direct encoding method yields the highest accuracy for both models, with LIF achieving 94.46\% and Lapicque achieving 93.23\%, while also demonstrating a moderate amount of sparsity with 92,069 spikes for LIF and 61,761 spikes for Lapicque. This indicates a good performance in terms of both accuracy and sparsity. In this case, the Lapicque model performs similarly to the LIF model, with slightly lower accuracy and fewer spikes. This makes it an attractive option for scenarios with stricter hardware requirements where minimizing the number of spikes is crucial.

In contrast, rate encoding shows moderate accuracy levels (77.99\% for LIF and 75.88\% for Lapicque) but results in much higher spike counts, particularly for the LIF model with 1,091,195 spikes, indicating lower sparsity. Delta encoding exhibits the lowest accuracy for both models (38.40\% for LIF and 40.53\% for Lapicque) and maintains the lowest spike counts (79,969 for LIF and 34,246 for Lapicque). Although delta encoding achieved the highest amount of sparsity, this was at the cost of a significant loss of accuracy.

In summary, the choice of neuron model and encoding method significantly impacts the accuracy and sparsity of SNNs. Direct encoding consistently outperforms rate and delta encoding in terms of accuracy, with the Lapicque neuron model exhibiting slightly lower accuracy but higher sparsity compared to the LIF model. While rate encoding offers moderate accuracy, it suffers from low sparsity. Delta encoding, despite achieving the highest sparsity, is impractical due to its significantly lower accuracy. The optimal choice depends on the specific application requirements, with direct encoding being preferable for high accuracy and the Lapicque model being advantageous for hardware-constrained scenarios prioritizing sparsity.

\section{Challenges of Hardware-Software Co-Design of SNNs}

While the potential benefits of SNNs are substantial, realizing these advantages in the real world necessitates careful co-design of the algorithms and the specialized hardware that supports them. In this section, we identify several key challenges that must be addressed to enable successful co-design for SNNs.

\subsubsection{Mapping algorithms to hardware} The complexity of the chosen neuron model has profound implications for hardware design.  Complex models (e.g., Hodgkin-Huxley) might demand a large number of computations per timestep \cite{arcas03_hodgkin_huxleyComputation}, straining embedded neuromorphic devices. Implementing on-chip learning rules, especially those beyond simple STDP, adds complexity in terms of memory technologies, update mechanisms, and potential trade-offs between flexibility and power consumption. Input encoding also plays a crucial role on the challenge of mapping SNN algorithms to hardware. For example, rate-based encoding can lead to dense activity, reducing the benefits of hardware-level sparsity support \cite{abderrahmane19_SNNHardware}, while temporal encoding might necessitate specialized hardware for spike-time processing \cite{oh20_hardwareSNN}.

\subsubsection{Scalability and network architectures} Building large-scale SNNs necessitates the efficient routing of the potentially massive number of spike events, requiring specialized routing fabrics or memory-centric architectures that reduce communication overhead \cite{carrillo11_SNN_routing}. Implementing diverse SNN topologies introduces unique challenges at both the software and hardware levels. For example, deep convolutional SNNs need efficient distribution of convolutional kernels and management of spike-based data \cite{aliyev24_pulse}. Replicating the connectivity of large-scale brain regions onto resource-constrained neuromorphic platforms might demand simplifying assumptions or distributed implementation strategies. Furthermore, real-time systems require optimized hardware-software mappings for real-time performance and sparsity \cite{meftah10_snnEdgeDetection}.

\subsubsection{Accuracy vs. efficiency trade-offs} SNN optimizations for efficiency, like reducing the bit precision of weights and activations (i.e., quantization) \cite{li22_quantizationSNN}, can significantly increase sparsity. However, such optimizations for efficiency might also carry the risk of severe accuracy degradation. Finding hardware-aware, optimal quantization strategies is important. Similarly, although pruning away weights creates sparsity, different SNN architectures might exhibit varying degrees of sensitivity to pruning. In addition, changes to the SNN architecture can create new trade-off considerations for different workloads. For instance, expanding an integrate-and-fire neuron model to a more bio-realistic leaky mechanism might increase the area overhead \cite{abderrahmane19_SNNHardware}, leading to important workload-specific trade-off considerations regarding the efficiency impacts of the more realistic neuron model.

\subsubsection{Neuromorphic hardware heterogeneity} Analog neuromorphic chips \cite{schemmel21_neuromorphicChip,miyashita17_neuromorphicChip} might offer superior energy efficiency but can suffer from device mismatch and noise, impacting accuracy. Digital platforms offer flexibility but could demand more complex circuitry to achieve equivalent sparsity benefits. The diversity of hardware also means navigating specialized programming tools and abstractions, potentially creating vendor lock-in and hindering the portability of SNN solutions.

\subsubsection{Lack of standardized tools and benchmarks} Comparing the performance of SNNs fairly across different algorithms and hardware platforms is hindered by a lack of standardization. Several current benchmarks focus on simple datasets (e.g., MNIST, FashionMNIST) \cite{sakemi23_snn}, which do not fully capture the strengths of SNNs in handling temporal or spiking sensor data (e.g., neuromorphic vision sensors, audio). While there is a growing body of work focused on developing neuromorphic datasets (e.g., \cite{he20_neuromorphicData}) the development of SNNs would benefit from more comprehensive benchmarks that include tasks like dynamic object recognition, spatiotemporal pattern analysis, and processing event-based sensor data, reflecting real-world scenarios where SNNs could excel.  Additionally, software frameworks for SNNs \cite{eshraghian21_snntorch} currently lack the maturity of debugging, profiling, and hardware mapping tools available for ANNs, potentially slowing down the research and deployment cycle of SNNs.

\section{Survey of Sparsity-Aware Hardware Architectures for SNNs}

\begin{table*}
\caption{Summary of hardware architectures for SNNs, with a focus on techniques for exploiting sparsity to enhance performance and energy efficiency}
\begin{tabular}{|p{2.5cm}|p{4cm}|p{5.5cm}|p{4cm}|}
\hline
\textbf{Work} & \textbf{Key Focus} & \textbf{Notable Features} & \textbf{Strengths} \\ \hline
Cerebron \cite{chen22_sparsityExploit_cerebron} & Spatiotemporal Sparsity Exploitation & Online workload scheduling, data reuse & Broad handling of sparsity, reduced computation time \\ \hline
MISS \cite{liu22_sparsityExploitation} & Irregular Sparsity Support & Unstructured pruning, sparsity-stationary dataflow & Efficient memory usage, handles complex sparsity patterns \\ \hline
ESSA \cite{kuang22_sparsityExploit_ESSA} & Inference Throughput Optimization & Adaptive spike compression, flexible fan-in/fan-out &  High throughput, handles both temporal \& spatial sparsity \\ \hline
SATA \cite{yin22_snn_sparsity_accel} & Sparsity-Aware Training & Systolic architecture, exploits sparsity in spikes, gradients, membrane potentials  & Improved training energy efficiency   \\ \hline
MF-DSNN \cite{zhang23_sparsityExploit} & Temporal Coding for Biological Realism & Multiplication-free coding, parallel neuron computations & Enhanced efficiency, aligns with biological mechanisms \\ \hline
Kuang et al. \cite{kuang22_sparsityExpoit} & Input Sparsity Handling & On-chip sparse weight storage, adaptive spike compression/decompression & Optimized for sparse inputs, high computational efficiency \\ \hline
SpinalFlow \cite{narayanan20_spinalflow} & Dataflow Optimization & Compressed time-stamped input spikes & Reduced storage overheads, lower energy consumption  \\ \hline
Aliyev et al. \cite{aliyev23_dse_snnAccelerators} & Design Space Exploration & Hardware \& model parameter alignment & Peak performance optimization, value of exploration \\ \hline
\end{tabular}
\label{tab:hardwareSurvey}
\end{table*}

Addressing the challenges in hardware-software co-design for sparsity-aware SNNs demands specialized hardware that can seamlessly handle the unique computational demands of these networks. The potential for extreme energy efficiency hinges on architectures explicitly designed to exploit irregular sparsity patterns and event-driven communication.  Table \ref{tab:hardwareSurvey} summarizes several innovative designs and approaches that have emerged. This section surveys some of the recent advances in this field, providing a representative overview of the strategies being explored to unlock the power of sparsity-aware SNN hardware.

One notable sparsity-aware implementation is Cerebron, a reconfigurable architecture that effectively handles both spatial and temporal sparsity in SNNs \cite{chen22_sparsityExploit_cerebron}. It utilizes an online channel-wise workload scheduling strategy to maximize data reuse and reduce computation time. This leads to significant reductions in prediction energy and faster processing, highlighting the importance of exploiting sparsity for neuromorphic computing.

Liu et al. \cite{liu22_sparsityExploitation} introduced the MISS (Memory-based Irregular Sparsity Support) framework to tackle irregular sparsity with a combination of software and hardware optimizations. The framework applies unstructured pruning to synaptic weights for increased efficiency.  The hardware utilizes a sparsity-stationary data flow to optimize memory usage and minimize processing overheads associated with sparsity, improving energy efficiency and speed. The MISS framework achieved an average of 36\% improvement in energy efficiency and 23\% speedup over baseline SNN accelerators by exploiting irregular sparsity in both input spikes and synaptic weights. Kuang et al. \cite{kuang22_sparsityExploit_ESSA} presented an accelerator called ESSA (Efficient Sparse SNN Accelerator) that targets both temporal sparsity (in spike events) and spatial sparsity (in weights) for enhanced SNN inference throughput. Key design features include adaptive spike compression for efficient handling of sparse spike patterns and a flexible fan-in/fan-out trade-off to work within neuromorphic system constraints. Results showed that ESSA achieved a performance equivalent of 253.1 GSOP/s and an energy efficiency of 32.1 GSOP/W for 75\% weight sparsity on a Xilinx Kintex Ultrascale FPGA, showing significant improvements in throughput and energy savings compared to other neuromorphic processors.

Unlike the prior works, which focused on inference, Yin et al. \cite{yin22_snn_sparsity_accel} proposed a sparsity-aware accelerator for training called SATA. SATA (Sparsity-Aware Training Accelerator for SNNs) focuses on making SNN training more efficient using backpropagation through time (BPTT).  Its systolic-based accelerator architecture exploits various forms of sparsity (in spikes, firing function gradients, and membrane potentials) to improve training energy efficiency.  SATA's analysis demonstrates that SNN training can be less energy-intensive than traditional ANN training. The analysis showed that although SNN training consumed approximately 1.27 times more total energy than ANNs when considering sparsity, it improved computational energy efficiency by 5.58 times over non-sparsity exploiting methods.

Another sparsity-aware implementation, the MF-DSNN accelerator \cite{zhang23_sparsityExploit}, focuses on a temporal coding scheme that removes the need for multiplication, enhancing biological realism. In concert with minimizing weight data access, this design achieves superior performance and energy efficiency, exemplifying the advantage of leveraging temporal sparsity.

Further exploring optimizations, an event-driven SNN accelerator by Kuang et al. \cite{kuang22_sparsityExpoit} features on-chip sparse weight storage and a self-adaptive spike compression/decompression mechanism, optimizing its handling of input spike sparsity. This enhances both speed and computational efficiency, evidenced by its high GSOPs/s performance even under elevated weight sparsity. The SpinalFlow architecture \cite{narayanan20_spinalflow} tackles SNN efficiency through a novel data flow strategy that processes compressed, time-stamped sequences of input spikes. This substantially reduces storage overheads and computational cost, demonstrating the power of optimized data handling in lowering energy consumption and boosting performance.

In the domain of sparsity-aware design space exploration, Aliyev et al. \cite{aliyev23_dse_snnAccelerators} focused on exploring the vast design space of sparsity-aware SNN accelerators. Their work sought configurations that provide peak performance by carefully aligning hardware and SNN model parameters. The proposed hardware leverages SNN sparsity for significant reductions in resource usage and increased speed. This work underscores the value of thorough design space exploration in creating highly efficient SNN accelerators. The proposed sparsity-aware SNN accelerator designs achieved up to 76\% reduction in hardware resources and a speed increase of up to 31.25 times, validating the effectiveness of tailoring hardware to specific sparsity-aware configurations for optimal performance.

Collectively, these advancements showcase the rapid progress in SNN accelerator design. By strategically exploiting different dimensions of sparsity, more efficient, effective, and biologically plausible computing models can be created.

\section{Conclusion}

We provided an overview of the hardware-software co-design of sparse SNNs, emphasizing the critical role of sparsity in achieving energy-efficient neuromorphic computing. Key takeaways include the understanding that sparsity is a dynamic property, influenced by various factors such as network architecture, training algorithms, neuron models, and input encoding methods. The exploration of different sparsity-aware hardware architectures reveals the potential for significant performance and energy efficiency gains through specialized designs that exploit irregular sparsity patterns and event-driven communications. The insights presented in this paper pave the way for future research in developing neuromorphic systems that fully exploit the computational advantages of sparse SNNs, enabling highly energy-efficient artificial intelligence in resource-constrained systems.

\section*{Acknowledgment}

This work was partially supported by the Technology and Research Initiative Fund (TRIF) provided to the University of Arizona by the Arizona Board of Regents (ABOR) and by the National Science Foundation under grant CNS-1844952.
	\balance
	\bibliographystyle{IEEEtran}
	\bibliography{tosiRefs,refs}

\begin{thebibliography}{10}
\providecommand{\url}[1]{#1}
\csname url@samestyle\endcsname
\providecommand{\newblock}{\relax}
\providecommand{\bibinfo}[2]{#2}
\providecommand{\BIBentrySTDinterwordspacing}{\spaceskip=0pt\relax}
\providecommand{\BIBentryALTinterwordstretchfactor}{4}
\providecommand{\BIBentryALTinterwordspacing}{\spaceskip=\fontdimen2\font plus
\BIBentryALTinterwordstretchfactor\fontdimen3\font minus \fontdimen4\font\relax}
\providecommand{\BIBforeignlanguage}[2]{{%
\expandafter\ifx\csname l@#1\endcsname\relax
\typeout{** WARNING: IEEEtran.bst: No hyphenation pattern has been}%
\typeout{** loaded for the language `#1'. Using the pattern for}%
\typeout{** the default language instead.}%
\else
\language=\csname l@#1\endcsname
\fi
#2}}
\providecommand{\BIBdecl}{\relax}
\BIBdecl

\bibitem{tavanaei19_deepSNN}
A.~Tavanaei, M.~Ghodrati, S.~R. Kheradpisheh, T.~Masquelier, and A.~Maida, ``Deep learning in spiking neural networks,'' \emph{Neural networks}, vol. 111, pp. 47--63, 2019.

\bibitem{foldiak03_sparseCoding}
P.~Foldiak, ``Sparse coding in the primate cortex,'' \emph{The handbook of brain theory and neural networks}, 2003.

\bibitem{spanne15_sparseCoding}
A.~Spanne and H.~J{\"o}rntell, ``Questioning the role of sparse coding in the brain,'' \emph{Trends in neurosciences}, vol.~38, no.~7, pp. 417--427, 2015.

\bibitem{faghihi22_sparseConnectivity}
F.~Faghihi, S.~Cai, and A.~A. Moustafa, ``A neuroscience-inspired spiking neural network for eeg-based auditory spatial attention detection,'' \emph{Neural Networks}, vol. 152, pp. 555--565, 2022.

\bibitem{eavani15_sparseConnect}
H.~Eavani, T.~D. Satterthwaite, R.~Filipovych, R.~E. Gur, R.~C. Gur, and C.~Davatzikos, ``Identifying sparse connectivity patterns in the brain using resting-state fmri,'' \emph{Neuroimage}, vol. 105, pp. 286--299, 2015.

\bibitem{brette12_computingSynchrony}
R.~Brette, ``Computing with neural synchrony,'' \emph{PLoS computational biology}, vol.~8, no.~6, p. e1002561, 2012.

\bibitem{yin22_snn_sparsity_accel}
R.~Yin, A.~Moitra, A.~Bhattacharjee, Y.~Kim, and P.~Panda, ``Sata: Sparsity-aware training accelerator for spiking neural networks,'' \emph{IEEE Transactions on Computer-Aided Design of Integrated Circuits and Systems}, 2022.

\bibitem{aliyev24_pulse}
I.~Aliyev and T.~Adegbija, ``Pulse: Parametric hardware units for low-power sparsity-aware convolution engine,'' \emph{arXiv preprint arXiv:2402.06210}, 2024.

\bibitem{ponulak11_snnIntro}
F.~Ponulak and A.~Kasinski, ``Introduction to spiking neural networks: Information processing, learning and applications,'' \emph{Acta neurobiologiae experimentalis}, vol.~71, no.~4, pp. 409--433, 2011.

\bibitem{burkitt06review}
A.~N. Burkitt, ``A review of the integrate-and-fire neuron model: I. homogeneous synaptic input,'' \emph{Biological cybernetics}, vol.~95, pp. 1--19, 2006.

\bibitem{lansky06_lif_neuron}
P.~Lansky, P.~Sanda, and J.~He, ``The parameters of the stochastic leaky integrate-and-fire neuronal model,'' \emph{Journal of Computational Neuroscience}, vol.~21, pp. 211--223, 2006.

\bibitem{nelson95_hodgkin}
M.~Nelson and J.~Rinzel, ``The hodgkin-huxley model,'' \emph{The book of genesis}, vol.~2, 1995.

\bibitem{she21_snnSpatioTemporal}
X.~She, S.~Dash, D.~Kim, and S.~Mukhopadhyay, ``A heterogeneous spiking neural network for unsupervised learning of spatiotemporal patterns,'' \emph{Frontiers in Neuroscience}, vol.~14, p. 615756, 2021.

\bibitem{serrano13_stdp}
T.~Serrano-Gotarredona, T.~Masquelier, T.~Prodromakis, G.~Indiveri, and B.~Linares-Barranco, ``Stdp and stdp variations with memristors for spiking neuromorphic learning systems,'' \emph{Frontiers in neuroscience}, vol.~7, p.~2, 2013.

\bibitem{han20_snnEnergyEfficiency}
B.~Han and K.~Roy, ``Deep spiking neural network: Energy efficiency through time based coding,'' in \emph{European Conference on Computer Vision}.\hskip 1em plus 0.5em minus 0.4em\relax Springer, 2020, pp. 388--404.

\bibitem{rojas96_backpropagation}
R.~Rojas and R.~Rojas, ``The backpropagation algorithm,'' \emph{Neural networks: a systematic introduction}, pp. 149--182, 1996.

\bibitem{bu23_ann_snn}
T.~Bu, W.~Fang, J.~Ding, P.~Dai, Z.~Yu, and T.~Huang, ``Optimal ann-snn conversion for high-accuracy and ultra-low-latency spiking neural networks,'' \emph{arXiv preprint arXiv:2303.04347}, 2023.

\bibitem{neftci19_surrogateGradients}
E.~O. Neftci, H.~Mostafa, and F.~Zenke, ``Surrogate gradient learning in spiking neural networks: Bringing the power of gradient-based optimization to spiking neural networks,'' \emph{IEEE Signal Processing Magazine}, vol.~36, no.~6, pp. 51--63, 2019.

\bibitem{rathi20_rateCoding}
N.~Rathi, G.~Srinivasan, P.~Panda, and K.~Roy, ``Enabling deep spiking neural networks with hybrid conversion and spike timing dependent backpropagation,'' \emph{arXiv preprint arXiv:2005.01807}, 2020.

\bibitem{kang2020asie}
Z.~Kang, L.~Wang, S.~Guo, R.~Gong, S.~Li, Y.~Deng, and W.~Xu, ``Asie: An asynchronous snn inference engine for aer events processing,'' \emph{ACM Journal on Emerging Technologies in Computing Systems (JETC)}, vol.~16, no.~4, pp. 1--22, 2020.

\bibitem{zhou21_temporalCoding}
S.~Zhou, X.~Li, Y.~Chen, S.~T. Chandrasekaran, and A.~Sanyal, ``Temporal-coded deep spiking neural network with easy training and robust performance,'' in \emph{Proceedings of the AAAI conference on artificial intelligence}, vol.~35, no.~12, 2021, pp. 11\,143--11\,151.

\bibitem{fellous04_spikePatterns}
J.-M. Fellous, P.~H. Tiesinga, P.~J. Thomas, and T.~J. Sejnowski, ``Discovering spike patterns in neuronal responses,'' \emph{Journal of Neuroscience}, vol.~24, no.~12, pp. 2989--3001, 2004.

\bibitem{sommer2022efficient}
J.~Sommer, M.~A. {\"O}zkan, O.~Keszocze, and J.~Teich, ``Efficient hardware acceleration of sparsely active convolutional spiking neural networks,'' \emph{IEEE Transactions on Computer-Aided Design of Integrated Circuits and Systems}, vol.~41, no.~11, pp. 3767--3778, 2022.

\bibitem{lew2022time}
D.~Lew, K.~Lee, and J.~Park, ``A time-to-first-spike coding and conversion aware training for energy-efficient deep spiking neural network processor design,'' in \emph{Proceedings of the 59th ACM/IEEE Design Automation Conference}, 2022, pp. 265--270.

\bibitem{yarga22_deltaEncoding}
S.~Y.~A. Yarga, J.~Rouat, and S.~Wood, ``Efficient spike encoding algorithms for neuromorphic speech recognition,'' in \emph{Proceedings of the International Conference on Neuromorphic Systems 2022}, 2022, pp. 1--8.

\bibitem{wang2022efficient}
Z.~Wang, X.~Gu, R.~S.~M. Goh, J.~T. Zhou, and T.~Luo, ``Efficient spiking neural networks with radix encoding,'' \emph{IEEE Transactions on Neural Networks and Learning Systems}, 2022.

\bibitem{wu19_snn_training}
Y.~Wu, L.~Deng, G.~Li, J.~Zhu, Y.~Xie, and L.~Shi, ``Direct training for spiking neural networks: Faster, larger, better,'' in \emph{Proceedings of the AAAI conference on artificial intelligence}, vol.~33, no.~01, 2019, pp. 1311--1318.

\bibitem{singh21_gestureSNN}
S.~Singh, A.~Sarma, S.~Lu, A.~Sengupta, V.~Narayanan, and C.~R. Das, ``Gesture-snn: Co-optimizing accuracy, latency and energy of snns for neuromorphic vision sensors,'' in \emph{2021 IEEE/ACM International Symposium on Low Power Electronics and Design (ISLPED)}.\hskip 1em plus 0.5em minus 0.4em\relax IEEE, 2021, pp. 1--6.

\bibitem{kim18_SNNPatternRecognition}
H.~Kim, S.~Hwang, J.~Park, S.~Yun, J.-H. Lee, and B.-G. Park, ``Spiking neural network using synaptic transistors and neuron circuits for pattern recognition with noisy images,'' \emph{IEEE Electron Device Letters}, vol.~39, no.~4, pp. 630--633, 2018.

\bibitem{amir17_dvsGesture}
A.~Amir, B.~Taba, D.~Berg, T.~Melano, J.~McKinstry, C.~Di~Nolfo, T.~Nayak, A.~Andreopoulos, G.~Garreau, M.~Mendoza, J.~Kusnitz, M.~Debole, S.~Esser, T.~Delbruck, M.~Flickner, and D.~Modha, ``A low power, fully event-based gesture recognition system,'' in \emph{Proceedings of the IEEE conference on computer vision and pattern recognition}, 2017, pp. 7243--7252.

\bibitem{yusob18_SNNanomalyDetection}
B.~Yusob, Z.~Mustaffa, and J.~Sulaiman, ``Anomaly detection in time series data using spiking neural network,'' \emph{Advanced Science Letters}, vol.~24, no.~10, pp. 7572--7576, 2018.

\bibitem{nichols12_snnRobot}
E.~Nichols, L.~J. McDaid, and N.~Siddique, ``Biologically inspired snn for robot control,'' \emph{IEEE transactions on cybernetics}, vol.~43, no.~1, pp. 115--128, 2012.

\bibitem{yu17_sparsityEnergy}
L.~Yu and Y.~Yu, ``Energy-efficient neural information processing in individual neurons and neuronal networks,'' \emph{Journal of Neuroscience Research}, vol.~95, no.~11, pp. 2253--2266, 2017.

\bibitem{bricken23_sparseNoise}
T.~Bricken, R.~Schaeffer, B.~Olshausen, and G.~Kreiman, ``Emergence of sparse representations from noise,'' 2023.

\bibitem{schweighofer01_sparseCoding}
N.~Schweighofer, K.~Doya, and F.~Lay, ``Unsupervised learning of granule cell sparse codes enhances cerebellar adaptive control,'' \emph{Neuroscience}, vol. 103, no.~1, pp. 35--50, 2001.

\bibitem{xu09_sparseCoding}
Y.~Xu, Z.~Xiao, and X.~Tian, ``A simulation study on neural ensemble sparse coding,'' in \emph{2009 International Conference on Information Engineering and Computer Science}.\hskip 1em plus 0.5em minus 0.4em\relax IEEE, 2009, pp. 1--4.

\bibitem{li16_robust_sparsity}
X.~Li, X.~Lu, and H.~Wang, ``Robust common spatial patterns with sparsity,'' \emph{Biomedical Signal Processing and Control}, vol.~26, pp. 52--57, 2016.

\bibitem{hassall18_sparsity_learning}
C.~D. Hassall, P.~C. Connor, T.~P. Trappenberg, J.~J. McDonald, and O.~E. Krigolson, ``Learning what matters: a neural explanation for the sparsity bias,'' \emph{International Journal of Psychophysiology}, vol. 127, pp. 62--72, 2018.

\bibitem{gerum20_sparsityPruning}
R.~C. Gerum, A.~Erpenbeck, P.~Krauss, and A.~Schilling, ``Sparsity through evolutionary pruning prevents neuronal networks from overfitting,'' \emph{Neural Networks}, vol. 128, pp. 305--312, 2020.

\bibitem{herbert22_sparsityModels}
E.~Herbert and S.~Ostojic, ``The impact of sparsity in low-rank recurrent neural networks,'' \emph{PLOS Computational Biology}, vol.~18, no.~8, p. e1010426, 2022.

\bibitem{harris19_sparsityModel}
K.~D. Harris, ``Additive function approximation in the brain,'' \emph{arXiv preprint arXiv:1909.02603}, 2019.

\bibitem{brunel07_lapicque}
N.~Brunel and M.~C. Van~Rossum, ``Lapicque’s 1907 paper: from frogs to integrate-and-fire,'' \emph{Biological cybernetics}, vol.~97, no.~5, pp. 337--339, 2007.

\bibitem{aliyev24_surrogateGradients}
I.~Aliyev and T.~Adegbija, ``Fine-tuning surrogate gradient learning for optimal hardware performance in spiking neural networks,'' \emph{arXiv preprint arXiv:2402.06211}, 2024.

\bibitem{eshraghian21_snntorch}
J.~K. Eshraghian, M.~Ward, E.~Neftci, X.~Wang, G.~Lenz, G.~Dwivedi, M.~Bennamoun, D.~S. Jeong, and W.~D. Lu, ``Training spiking neural networks using lessons from deep learning,'' \emph{arXiv preprint arXiv:2109.12894}, 2021.

\bibitem{dong2017learning}
Y.~Dong, R.~Ni, J.~Li, Y.~Chen, J.~Zhu, and H.~Su, ``Learning accurate low-bit deep neural networks with stochastic quantization,'' \emph{arXiv preprint arXiv:1708.01001}, 2017.

\bibitem{arcas03_hodgkin_huxleyComputation}
B.~A. y~Arcas, A.~L. Fairhall, and W.~Bialek, ``Computation in a single neuron: Hodgkin and huxley revisited,'' \emph{Neural computation}, vol.~15, no.~8, pp. 1715--1749, 2003.

\bibitem{abderrahmane19_SNNHardware}
N.~Abderrahmane and B.~Miramond, ``Information coding and hardware architecture of spiking neural networks,'' in \emph{2019 22nd Euromicro Conference on Digital System Design (DSD)}.\hskip 1em plus 0.5em minus 0.4em\relax IEEE, 2019, pp. 291--298.

\bibitem{oh20_hardwareSNN}
S.~Oh, D.~Kwon, G.~Yeom, W.-M. Kang, S.~Lee, S.~Y. Woo, J.~S. Kim, M.~K. Park, and J.-H. Lee, ``Hardware implementation of spiking neural networks using time-to-first-spike encoding,'' \emph{arXiv preprint arXiv:2006.05033}, 2020.

\bibitem{carrillo11_SNN_routing}
S.~Carrillo, J.~Harkin, L.~McDaid, S.~Pande, S.~Cawley, and F.~Morgan, ``Adaptive routing strategies for large scale spiking neural network hardware implementations,'' in \emph{Artificial Neural Networks and Machine Learning--ICANN 2011: 21st International Conference on Artificial Neural Networks, Espoo, Finland, June 14-17, 2011, Proceedings, Part I 21}.\hskip 1em plus 0.5em minus 0.4em\relax Springer, 2011, pp. 77--84.

\bibitem{meftah10_snnEdgeDetection}
B.~Meftah, O.~Lezoray, and A.~Benyettou, ``Segmentation and edge detection based on spiking neural network model,'' \emph{Neural Processing Letters}, vol.~32, pp. 131--146, 2010.

\bibitem{li22_quantizationSNN}
C.~Li, L.~Ma, and S.~Furber, ``Quantization framework for fast spiking neural networks,'' \emph{Frontiers in Neuroscience}, vol.~16, p. 918793, 2022.

\bibitem{schemmel21_neuromorphicChip}
J.~Schemmel, S.~Billaudelle, P.~Dauer, and J.~Weis, ``Accelerated analog neuromorphic computing,'' in \emph{Analog Circuits for Machine Learning, Current/Voltage/Temperature Sensors, and High-speed Communication: Advances in Analog Circuit Design 2021}.\hskip 1em plus 0.5em minus 0.4em\relax Springer, 2021, pp. 83--102.

\bibitem{miyashita17_neuromorphicChip}
D.~Miyashita, S.~Kousai, T.~Suzuki, and J.~Deguchi, ``A neuromorphic chip optimized for deep learning and cmos technology with time-domain analog and digital mixed-signal processing,'' \emph{IEEE Journal of Solid-State Circuits}, vol.~52, no.~10, pp. 2679--2689, 2017.

\bibitem{sakemi23_snn}
Y.~Sakemi, K.~Yamamoto, T.~Hosomi, and K.~Aihara, ``Sparse-firing regularization methods for spiking neural networks with time-to-first-spike coding,'' \emph{Scientific Reports}, vol.~13, no.~1, p. 22897, 2023.

\bibitem{he20_neuromorphicData}
W.~He, Y.~Wu, L.~Deng, G.~Li, H.~Wang, Y.~Tian, W.~Ding, W.~Wang, and Y.~Xie, ``Comparing snns and rnns on neuromorphic vision datasets: Similarities and differences,'' \emph{Neural Networks}, vol. 132, pp. 108--120, 2020.

\bibitem{chen22_sparsityExploit_cerebron}
Q.~Chen, C.~Gao, and Y.~Fu, ``Cerebron: A reconfigurable architecture for spatiotemporal sparse spiking neural networks,'' \emph{IEEE Transactions on Very Large Scale Integration (VLSI) Systems}, vol.~30, no.~10, pp. 1425--1437, 2022.

\bibitem{liu22_sparsityExploitation}
F.~Liu, Z.~Wang, W.~Zhao, Y.~Chen, T.~Yang, X.~Yang, and L.~Jiang, ``Randomize and match: Exploiting irregular sparsity for energy efficient processing in snns,'' in \emph{2022 IEEE 40th International Conference on Computer Design (ICCD)}.\hskip 1em plus 0.5em minus 0.4em\relax IEEE, 2022, pp. 451--454.

\bibitem{kuang22_sparsityExploit_ESSA}
Y.~Kuang, X.~Cui, Z.~Wang, C.~Zou, Y.~Zhong, K.~Liu, Z.~Dai, D.~Yu, Y.~Wang, and R.~Huang, ``Essa: Design of a programmable efficient sparse spiking neural network accelerator,'' \emph{IEEE Transactions on Very Large Scale Integration (VLSI) Systems}, vol.~30, no.~11, pp. 1631--1641, 2022.

\bibitem{zhang23_sparsityExploit}
Y.~Zhang, S.~Wang, and Y.~Kang, ``Mf-dsnn: An energy-efficient high-performance multiplication-free deep spiking neural network accelerator,'' in \emph{2023 IEEE 5th International Conference on Artificial Intelligence Circuits and Systems (AICAS)}.\hskip 1em plus 0.5em minus 0.4em\relax IEEE, 2023, pp. 1--4.

\bibitem{kuang22_sparsityExpoit}
Y.~Kuang, X.~Cui, C.~Zou, Y.~Zhong, Z.~Dai, Z.~Wang, K.~Liu, D.~Yu, and Y.~Wang, ``An event-driven spiking neural network accelerator with on-chip sparse weight,'' in \emph{2022 IEEE International Symposium on Circuits and Systems (ISCAS)}.\hskip 1em plus 0.5em minus 0.4em\relax IEEE, 2022, pp. 3468--3472.

\bibitem{narayanan20_spinalflow}
S.~Narayanan, K.~Taht, R.~Balasubramonian, E.~Giacomin, and P.-E. Gaillardon, ``Spinalflow: An architecture and dataflow tailored for spiking neural networks,'' in \emph{2020 ACM/IEEE 47th Annual International Symposium on Computer Architecture (ISCA)}.\hskip 1em plus 0.5em minus 0.4em\relax IEEE, 2020, pp. 349--362.

\bibitem{aliyev23_dse_snnAccelerators}
I.~Aliyev, K.~Svoboda, and T.~Adegbija, ``Design space exploration of sparsity-aware application-specific spiking neural network accelerators,'' \emph{IEEE Journal on Emerging and Selected Topics in Circuits and Systems}, 2023.

\end{thebibliography}
	
\end{document}